\documentclass[aps,twocolumn,amsmath,amssymb,showpacs]{revtex4-1}
\usepackage[ansinew]{inputenc}
\usepackage{graphicx}
\usepackage{xcolor}
\usepackage{siunitx}
\usepackage{hyperref}

\begin{document}

\title{Narrow optical transitions in erbium-implanted silicon waveguides}
\author{Andreas Gritsch}
\email{A.G. and L.W. contributed equally to this work.}
\author{Lorenz Weiss}
\email{A.G. and L.W. contributed equally to this work.}
\author{Johannes Fr\"uh}
\author{Stephan Rinner}
\author{Andreas Reiserer}
\email{andreas.reiserer@tum.de}

\affiliation{Max-Planck-Institut f\"ur Quantenoptik, Quantum Networks Group, Hans-Kopfermann-Strasse 1, D-85748 Garching, Germany}
\affiliation{Technical University of Munich, TUM School of Natural Sciences and Munich Center for Quantum Science and Technology (MCQST), James-Franck-Straße 1, D-85748 Garching, Germany }

\begin{abstract}
The realization of a scalable architecture for quantum information processing is a major challenge of quantum science. A promising approach is based on emitters in nanostructures that are coupled by light. Here, we show that erbium dopants can be reproducibly integrated at well-defined lattice sites by implantation into pure silicon. We thus achieve a narrow inhomogeneous broadening, $<1\,\si{\giga\hertz}$, strong optical transitions, and an outstanding optical coherence even at temperatures of $8\,\si{\kelvin}$, with an upper bound to the homogeneous linewidth of  $\sim10\,\si{\kilo\hertz}$. Our study thus introduces a promising materials platform for the implementation of on-chip quantum memories, microwave-to-optical conversion, and distributed quantum information processing.
\end{abstract}

\maketitle

\section{Introduction} \label{sec:Intro}

Individual dopants in solids are prime candidates for the realization of quantum computers \cite{vandersypen_interfacing_2017} and quantum networks \cite{reiserer_cavity-based_2015, ruf_quantum_2021}. While a large variety of host materials is explored \cite{wolfowicz_quantum_2021}, the use of silicon seems particularly attractive: First, nanofabrication has reached a unique level of maturity in this material. Second, a platform for integrated photonic processing, both classical \cite{vivien_handbook_2013} and quantum \cite{arrazola_quantum_2021}, is readily available. Finally, defect-free and isotopically purified crystalline layers can be obtained on a wafer-scale \cite{mazzocchi_99992_2019}, which can largely eliminate the decoherence via nuclear spins and paramagnetic impurities \cite{saeedi_room-temperature_2013} and reduce the inhomogeneous broadening of optical transitions \cite{chartrand_highly_2018}.

To harness the mentioned advantages for quantum applications, different emitters in silicon have been explored \cite{morse_photonic_2017, chartrand_highly_2018, weiss_erbium_2021, durand_broad_2021}, and single dopants have been detected optically \cite{yin_optical_2013, redjem_single_2020, higginbottom_optical_2022}. Among these emitters, and also those in all other host materials \cite{wolfowicz_quantum_2021}, erbium stands out since it offers an optical transition in the main wavelength band of telecommunication, where loss in optical fibers is minimal. This transition exhibits exceptional coherence in several hosts \cite{liu_spectroscopic_2005, bottger_optical_2006}, approaching the lifetime-limit in resonators with sufficient Purcell enhancement \cite{merkel_coherent_2020}. Here, the stability and narrow width of the optical lines also facilitate spectrally multiplexed control \cite{chen_parallel_2020, ulanowski_spectral_2022} of many erbium dopants within the same resonator. This eliminates the challenging requirement of single-defect positioning encountered with other emitters \cite{ruf_quantum_2021}.

In spite of the promise of integrating erbium dopants into nanophotonic silicon structures, a clear and reproducible path to an understood site with good properties has been lacking. Many previous experiments, mainly in the context of lasers, studied erbium fluorescence after off-resonant excitation \cite{vinh_photonic_2009, kenyon_erbium_2005}. These experiments only observed a small fraction of the implanted erbium dopants --- those that can capture electrons from the conduction band and then decay radiatively. Our recent work \cite{weiss_erbium_2021} cut through this by measuring the fluorescence after resonant excitation, which allows one to observe and characterize erbium in all sites that decay radiatively. Instead of a wide distribution with numerous peaks observed in all earlier studies with implanted silicon \cite{przybylinska_optically_1996, kenyon_erbium_2005}, we found a rather limited number of narrow lines. Still, their number and position depended on the sample preparation, in particular the annealing conditions. At first sight, this is surprising, as silicon is a monoatomic crystal in which all atoms are identically coordinated. Thus, one would not expect that a single dopant can be integrated in many different ways --- unless it is gettering, i.e. it tends to form clusters with other dopants and impurities. In this case, the higher the implantation dose, annealing temperature, and impurity concentration of the starting material, the more different clusters can form and the more optical lines are expected. This explains the numerous peaks found in previous studies \cite{przybylinska_optically_1996, kenyon_erbium_2005, berkman_sub-megahertz_2021, weiss_erbium_2021}.

However, quantum applications favor dopant integration at a single, well-defined, and reproducible site with a high yield. In this work, we achieve this by starting from silicon layers with a low concentration of impurities, and  avoiding annealing at temperatures exceeding $\SI{800}{\kelvin}$. In several samples of different type, this allows us to integrate erbium at the same, previously unobserved sites that exhibit very favorable properties. In particular, we observe the fastest optical transitions of erbium dopants in any known host material, and exceptionally large crystal-field splittings, which leads to low Orbach decoherence and thus enables narrow homogeneous linewidths $<10\,\si{\kilo\hertz}$ up to a temperature of $8\,\si{\kelvin}$, which is conveniently accessible with dry $^4\text{He}$ cryocoolers. These improvements pave the way for using erbium-implanted silicon for quantum applications.

\section{Samples and experimental setup} \label{sec:SamplesAndExperimentalSetup}

\begin{figure}[tb]
\includegraphics[width=1.0\columnwidth]{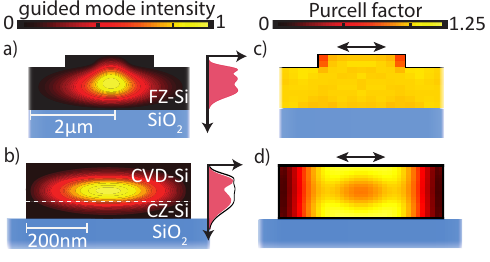}
\caption{        \label{fig:Waveguides}
\textbf{Different waveguide geometries used in this work.} a) To confine a single guided mode (colors: intensity profile), rib waveguides are fabricated by partially etching a chip with a $\SI{2}{\micro\meter}$ FZ ($>\SI{10}{\kilo \ohm \centi\meter}$) device layer (DL) on top of a $\SI{1}{\micro\meter}$ buried oxide (BOX). Erbium is implanted at three different energies to give an approximately homogeneous  concentration in the waveguide (red profile on the right). b) Ridge waveguides are fabricated by fully etching a $\SI{0.2}{\micro\meter}$ thin DL on a $>\SI{2}{\micro\meter}$ BOX. The intensity profile (colors) shows the fundamental TE mode. The DL consists of CZ material ($>\SI{1}{\ohm \centi\meter}$), or of a $\SI{0.12}{\micro\meter}$ pure CVD layer grown on top of a $\SI{0.07}{\micro\meter}$ thin CZ seed. To achieve an approximately homogeneous concentration (right), again three (CZ, black) or two (CVD, red) different implantation energies are used. c) In the rib waveguides, the photonic local density of states and the associated Purcell factor in the silicon layer (colors) is independent of the dopant position in the waveguide, leading to an effective Purcell factor $F_P$ of 1 for all emitters. d) In the ridge waveguides, the radiative decay rate of dipolar emitters oriented along the waveguide width (double arrow) is suppressed ($F_P < 1$) or enhanced, depending on its position.}
\end{figure}

To investigate the integration of erbium and its dependence on the silicon purity, we prepared three different types of samples, as shown in Fig.~\ref{fig:Waveguides}. We label them according to the crystal growth technique used for the erbium-containing silicon layer: Float-zone -- "FZ", Czochralski -- "CZ", and chemical vapor deposition -- "CVD".  The fabrication of all samples starts from commercial silicon-on-insulator (SOI) wafers from different suppliers made by the bond-and-cut-back (FZ) or smart-cut (CZ, CVD) technique. The CVD samples involve an additional growth step: After exposing them to a hydrogen bake at $\sim \SI{1200}{\kelvin}$ to prepare a clean silicon surface, an epitaxial CVD layer of pure silicon with natural isotope abundance is grown to a total device layer thickness of $0.19\,\si{\micro\meter}$. More information about the used silicon layers is given in Appendix \ref{app:SamplePreparation}.

The samples are implanted with erbium at different facilities using different energy and dose to achieve an approximately homogeneous peak concentration of $2\cdot 10^{16}\,\si{\centi\meter}^{-3}$ (FZ) and $2 \cdot 10^{17}\,\si{\centi\meter}^{-3}$ (CVD and CZ). This moderate value is chosen to avoid large waveguide loss from implantation-induced damage \cite{gad_loss_2003}, and amplified spontaneous emission that would complicate lifetime measurements. In contrast to previous works \cite{gad_loss_2003, kenyon_erbium_2005, yin_optical_2013, weiss_erbium_2021}, no oxygen implantation and no post-implantation thermal treatment exceeding $\sim 800\,\si{\kelvin}$ is applied. This temperature is a compromise, intended to be hot enough to largely repair the implantation damage, but cold enough to prevent a significant diffusion of the erbium dopants that would lead to segregation and undesired cluster sites \cite{kenyon_erbium_2005}.

After the implantation, either nanophotonic rib (on the FZ sample) or ridge waveguides (on the CZ and CVD samples) are fabricated by electron-beam lithography and reactive ion etching in fluorine chemistry, as shown in Fig.~\ref{fig:Waveguides} and detailed in Appendix \ref{app:WvgFabrication}. In the CVD samples, we measure that the implantation increases the waveguide loss from $(3.5 \pm 1.2)\,\text{dB/cm}$ to $(5.7 \pm 1.2)\,\text{dB/cm}$, similar to previous work \cite{gad_loss_2003}. This value may be further optimized in a systematic study of the implantation and annealing conditions.

\begin{figure*}[tb]
\includegraphics[width=2.0\columnwidth]{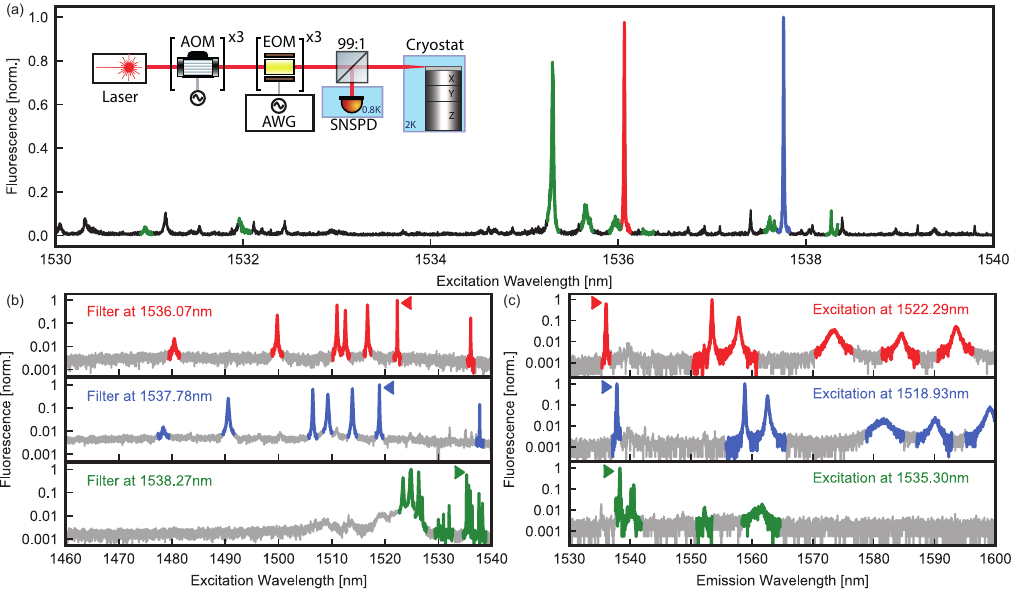}
\caption{        \label{fig:CrystalFieldLevels}
\textbf{Spectroscopy of erbium-implanted silicon waveguides.} (a) Inset: Experimental setup. A laser is switched by acousto-optic (AOM) and phase-modulated by electro-optic modulators (EOM). A 99:1 beam splitter is used to guide the light emitted from the sample in the cryostat to single-photon detectors (SNSPD). Main graph: Scanning the excitation laser frequency reveals several narrow fluorescence peaks, here shown for the FZ sample. (b) Insertion of a narrowband optical filter, tuned on resonance with one of the three most prominent peaks (color coding), eliminates a broadband background observed otherwise (which is strongest in the shown CVD samples). Scanning the excitation laser then reveals the excited state crystal fields of the individual sites. (c) Keeping the laser on resonance with one of the main peaks while scanning the filter gives the ground state crystal fields. The filled triangles in c(b) mark the filter (excitation laser) position in b(c).
}
\end{figure*}

To achieve a strong signal in spite of the moderate erbium concentration, we use up to several $\si{\milli\meter}$ long waveguides, which are coupled to single-mode optical fibers using butt-coupling (FZ) or tapered-fiber-coupling (CZ and CVD) \cite{vivien_handbook_2013}. The samples are mounted in a closed-cycle cryostat at adjustable temperature between 2 and $\SI{20}{\kelvin}$. Using the setup shown in the inset of Fig.~\ref{fig:CrystalFieldLevels}a, resonant fluorescence spectroscopy reveals the wavelengths at which the implanted dopants can be excited.

The transitions between the lowest crystal field levels of the inner 4f-shell of erbium occur between $1530$ and $1540\,\si{\nano\meter}$ \cite{liu_spectroscopic_2005}. When using a pulsed scheme to separate the emission from reflections of the excitation laser, we find several narrow emission lines in this range, as shown in Fig.~\ref{fig:CrystalFieldLevels}a for the FZ samples and in Appendix \ref{app:FluorescenceSpectrum} for the others. The former exhibit a smaller number of resonant peaks, which is attributed to the higher purity of the starting material. Still, in spite of the differing material and implantation conditions, we find the same three major lines in all studied sample types. Their relative intensity is slightly changed. Taken together, this indicates that in pure silicon, erbium can be reproducibly integrated at three main sites. Compared to our previous work that used higher annealing temperature \cite{weiss_erbium_2021}, the wavelength of the peaks is changed (see Appendix \ref{app:FluorescenceSpectrum}). The shift is too large to be caused by strain \cite{zhang_single_2019}, indicating erbium integration at different sites. In the following, we show that two of the sites discovered in this work - those with the strongest signal - have very favorable properties for quantum applications.

\section{Crystal field measurements} \label{sec:CrystalFieldMeasurements}

In rare-earth doped crystals, the transition strength and optical coherence depend on the splitting of the 4f levels by the crystal field (CF) \cite{liu_spectroscopic_2005}. To determine this splitting, we insert a tunable and narrowband optical filter into the detection path. We first keep the filter resonant with one of the brightest peaks in the above measurement and scan the excitation laser to shorter wavelengths. As can be seen in Fig.~\ref{fig:CrystalFieldLevels}b, we can thus identify the higher lying CF levels in the optically excited state manifold. We then keep the laser at the brightest peak and scan the filter frequency to determine the energy levels in the ground state CF manifold, see Fig.~\ref{fig:CrystalFieldLevels}c. The high signal-to-noise ratio enabled by our low-loss waveguides ensures that all levels that are in reach of the laser and filter are observed. The broadening of the long-wavelength peaks indicates a fast phononic relaxation of the upper CF levels, as commonly observed with rare-earth dopants \cite{liu_spectroscopic_2005}. Assigning the observed peaks, we again find that in all sample types, erbium is integrated at three dominant sites, which we label "A" (blue), "B" (red) and "P" (green).

We also observe an additional site labelled "O", which is "optically active" \cite{kenyon_erbium_2005} (not marked). This means that it can be excited by electrons in the conduction band. In agreement with earlier studies, only a small fraction of all erbium dopants is integrated in this site. Still, it leads to fluorescence that is independent of the excitation wavelength provided the power is large enough to give a significant above-bandgap excitation (see Appendix \ref{app:PandOSites}). While coupling to the conduction band is key to the realization of electrically-driven photon emission and lasing \cite{kenyon_erbium_2005, vinh_photonic_2009}, it is not required for quantum applications, and may even be detrimental to the optical coherence. Therefore, we focus on the three bright sites which are not coupled to the conduction band. One of them, site "P" (green), shows a two-fold larger number of crystal fields than possible for single erbium dopants in the absence of a magnetic field. We therefore attribute these lines to erbium pairs or precipitates. The mutual coupling of emitters in these sites may lead to coherence times that are too short for quantum applications.

In contrast, we observe only seven CF levels in the excited state for the two other main sites "A" (blue, $1537.761(2)\,\si{\nano\meter}$) and "B" (red, $1536.061(2)\,\si{\nano\meter}$), as expected for erbium dopants in low-symmetry sites \cite{liu_spectroscopic_2005}. We observe a large splitting between the lowest two CF levels in both the ground state (site A: $\SI{2634(1)}{\giga\hertz}$, site B: $\SI{2187(1)}{\giga\hertz}$) and in the excited state (site A: $\SI{2418(1)}{\giga\hertz}$, site B:  $\SI{1766(1)}{\giga\hertz}$). This is considerably larger than in all previously investigated erbium hosts, including $\mathrm{Y}_2\mathrm{SiO}_5$, $\mathrm{Y}_2\mathrm{O}_3$, $\mathrm{Y}_3\mathrm{Al}_5\mathrm{O}_12$, $\mathrm{LiNbO}_3$, $\mathrm{YVO}_4$ and $\mathrm{CaWO}_4$ (see Appendix \ref{app:OtherHostMaterials}). Combined with the higher Debye temperature of silicon \cite{wolfowicz_quantum_2021}, this can improve the optical coherence at higher temperature, as we will show below.

\begin{figure}[tb]
\includegraphics[width=1.\columnwidth]{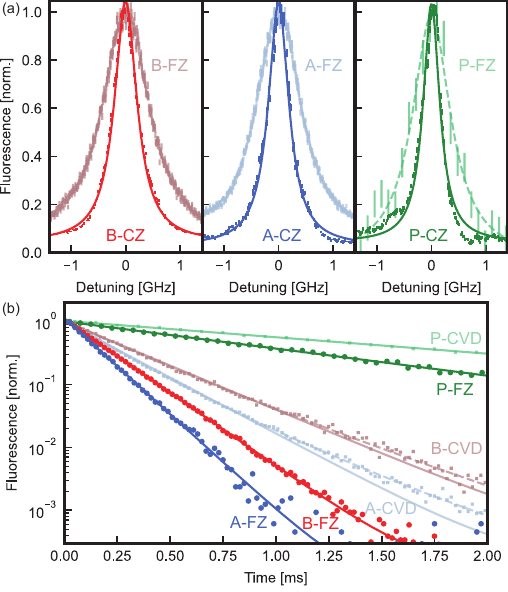}
\caption{ \label{fig:InhomLine_Purcell_Lifetime}
\textbf{Inhomogeneous linewidth and optical lifetime}. a) Resonant fluorescence after excitation laser pulses of different frequency. The observed lines are well-fit by Lorentzian curves. The observed linewidths in the FZ samples (faint colors) are larger than in the CZ (bright colors) and CVD samples (see Appendix \ref{app:LineWidth}) because of implantation-induced damage.  b) To determine the lifetime, we excite a higher CF level and observe the filtered lowest CF transition. The observed lifetime depends on the site (color) and the waveguide geometry (faint: ridge waveguide, bright: rib waveguide). The slowest decay is observed for site P (green), whereas site A (blue) and B (red) show faster decays. In the ridge waveguides, the spatial variation of the local density of states leads to a slower bi-exponential (dashed) rather than a single-exponential (solid) decay (see Appendix \ref{app:LDOS}).
}
\end{figure}

In addition to the favorable CF splitting, the optical transitions of the newly discovered sites A and B also exhibit a large branching ratio, defined as the fraction of photons emitted on the transition between the lowest CF levels in the ground and optically excited state manifold. This transition is most relevant for quantum applications, as the involved levels cannot relax via phonon emission \cite{liu_spectroscopic_2005}. By measuring the light emitted into the FZ waveguides with and without the narrowband filter, and after calibrating the filter transmission, we determine a branching ratio of $23(5)\,\%$ for both site A and B, comparable to other erbium hosts (see Appendix \ref{app:OtherHostMaterials}) and to the Debye-Waller factor of other emitters in silicon \cite{bergeron_silicon-integrated_2020}. The uncertainty estimate reflects the wavelength dependence of the detector efficiency and the fiber-optical elements that lead to a reduced detection probability for the emission into the highest-lying CF levels. In addition, the measurement may have a small systematic deviation caused by different optical dipole orientations of the involved transitions that can entail different waveguide couplings.

\section{Inhomogeneous broadening} \label{sec:InhomogeneousBroadening}

For all measured sites, we observe narrow inhomogeneous lines $\lesssim \SI{1}{\giga\hertz}$ in all samples, as shown in Fig.~\ref{fig:InhomLine_Purcell_Lifetime}a. The narrowest lines are obtained for the CZ sample, with a full-width-at-half-maximum (FWHM) of $\SI{0.37(2)}{\giga\hertz}$ (site P, green), $\SI{0.46(1)}{\giga\hertz}$ (site A, blue), and $\SI{0.50(1)}{\giga\hertz}$ (site B, red), comparable to or even narrower than in other hosts and in our earlier work \cite{weiss_erbium_2021}. In the CVD samples, we observe two peaks with comparable FWHM (see Appendix \ref{app:LineWidth}). In the FZ implantation, we observe significantly broader lines of $\SI{0.96(1)}{\giga\hertz}$ (site A), $\SI{1.07(1)}{\giga\hertz}$ (site B), and $\SI{0.86(6)}{\giga\hertz}$ (site P). Since the erbium concentration in these samples is tenfold smaller than in the others, we attribute the increased inhomogeneous broadening to a more pronounced damage of the crystalline layer caused by the higher implantation energies required in thicker waveguides.

The observation of such narrow resonances in all samples indicates that the dopants are integrated at well-defined lattice sites without inducing a large strain inhomogeneity at the studied concentrations. Also, the implantation and annealing does not leave a substantial damage to the crystal structure, which would be detrimental to the coherence time of embedded rare-earth dopants \cite{liu_spectroscopic_2005}. Thus, these results constitute a key step towards quantum-controlled applications of erbium-doped silicon, in particular frequency-multiplexed single emitters \cite{ulanowski_spectral_2022, chen_parallel_2020}, frequency converters \cite{obrien_interfacing_2014} and on-chip quantum memories \cite{craiciu_multifunctional_2021} that require narrow lines to achieve sufficient optical depth.

Similar to other host crystals, the remaining inhomogeneous broadening is attributed to random strain fields in the waveguide that are caused by the different thermal expansion coefficients of Si and $\mathrm{SiO}_2$, by small residual amounts of oxygen and other impurities, and finally by the mixed isotopic composition of the crystal. We expect that similar to other emitters in silicon, these sources can be largely eliminated in optimized samples \cite{chartrand_highly_2018, liu_28silicon--insulator_2022}, opening exciting prospects for ultra-narrow optical lines in nanofabricated waveguide and resonator structures.

\section{Optical lifetime} \label{sec:OpticalLifetime}

In addition to the large CF splitting and the narrow inhomogeneous broadening, the sites discovered in this work have another advantage: They exhibit comparably fast optical transitions. The temporal decay of the fluorescence after the excitation laser is switched off is shown in Fig.~\ref{fig:InhomLine_Purcell_Lifetime}b. In the FZ samples, the newly-observed sites A and B exhibit a decay constant of $\SI{142(1)}{\micro\second}$ and $\SI{186(1)}{\micro\second}$, respectively, about tenfold faster than previously observed sites in silicon \cite{kenyon_erbium_2005, vinh_photonic_2009, weiss_erbium_2021}. This may indicate that the dopants are located in interstitial sites, in which the local-field correction factor in materials with a high refractive index, such as silicon, can lead to accelerated spontaneous emission  \cite{de_vries_resonant_1998}. The measured large CF splittings may strengthen this hypothesis, but further experimental investigation is needed for a definite statement.

In principle, also non-radiative decay processes may contribute to the short lifetime. However, in contrast to experiments that investigate erbium sites that are coupled to the conduction band \cite{yin_optical_2013, hu_time-resolved_2022}, the lifetime in our measurements does not depend on temperature (at least up to $\SI{20}{\kelvin}$, see below), excitation power and excitation pulse duration. In addition,  the lifetime strongly depends on the optical density of states, which changes with the geometry of the waveguide. Thus, in the CVD and CZ samples (faint colors in Fig.~\ref{fig:InhomLine_Purcell_Lifetime}b), a slower decay is observed for all three sites, as the Purcell factor in the ridge waveguides at the erbium position is often smaller than one, c.f. Fig.~\ref{fig:Waveguides}c and d, which leads to a suppression of the emission. Further analysis is provided in Appendix \ref{app:LDOS}. Comparing the decay rate in the different samples types, we infer that the fast decay of the newly discovered sites is predominantly radiative. 

When using single dopants as qubits, a fast radiative decay results in a significant speedup of single-photon generation and other operations, e.g.~almost 100-fold when compared to the commonly used host material YSO \cite{chen_parallel_2020, merkel_coherent_2020}. If desired, the lifetime can be further reduced by integrating the erbium emitters into Fabry-Perot \cite{merkel_coherent_2020, ulanowski_spectral_2022} or photonic crystal \cite{chen_parallel_2020} resonators. But even in the absence of such enhancement, the observed decay times of sites A and B are faster than the time it takes light to travel $100\,\si{\kilo\meter}$ along an optical fiber. Thus, the entanglement rate in global quantum networks based on individual emitters would not be limited by the photon temporal length, but by the signaling time. 

\section{Homogeneous linewidth} \label{sec:HomogeneousBroadening}

\begin{figure}[tb]
\includegraphics[width=1.0\columnwidth]{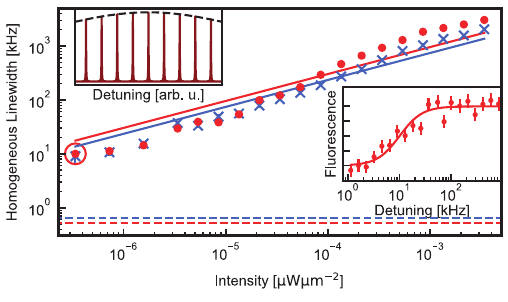}
\caption{\label{fig:PowerDependence}
\textbf{Homogeneous linewidth measurement.} Top left inset: Measurement scheme. The laser is modulated to generate a comb of up to 27 lines with equal separation and amplitude, exciting many subensembles (brown) within the inhomogeneous line of the dopants (dashed). When the modulation frequency is smaller than the linewidth of the subensembles, saturation leads to a signal decrease (right inset), which is well-fit by Lorentzian curves (red solid) with FWHM down to $18(6)\,\si{\kilo\hertz}$ for site A (blue), and $20(4)\,\si{\kilo\hertz}$ for site B (red). An upper bound for the homogeneous linewidth is given by half the width of the transient hole (main graph, CVD sample, red and blue data). The data exhibit a $\sqrt{I}$ scaling (lines) down to the lowest intensity that provides sufficient signal-to-noise (red circle, shown as inset). This indicates that the measured linewidth is determined by power broadening and thus constitutes an upper bound to the homogeneous linewidth of erbium dopants in the investigated nanophotonic silicon waveguides.}

\end{figure}

To use erbium-doped silicon for quantum applications, the coherence of the optical transition is paramount \cite{ruf_quantum_2021}. We therefore characterize the homogeneous linewidth via transient spectral hole burning \cite{szabo_observation_1975, volker_hole-burning_1989}. Our measurement technique \cite{weiss_erbium_2021} is based on the observation that the fluorescence signal $S$ increases nonlinearly with the laser intensity $I$ applied at a single frequency, $S \propto \sqrt{I}$, because of saturation. We therefore modulate the input laser field with three concatenated electro-optical modulators in order to apply up to 27 laser fields of about the same intensity $I$. The fields have an equidistant frequency separation within the inhomogeneous linewidth, exciting different erbium subensembles as shown in the top left inset of Fig.~\ref{fig:PowerDependence}.

If the detuning of the laser fields is larger than the ensemble linewidth, their simultaneous irradiation will lead to a 27-fold increase of the fluorescence compared to that of a single field. If, however, the detuning is small, or zero, the fluorescence will only increase by a factor of $\sim \sqrt{27}$ (see Appendix \ref{app:HomLinewidth}). Thus, using excitation pulses of $0.15\, \si{\milli\second}$ duration and scanning the modulation frequency, we can measure the ensemble linewidth on the timescale of the pulses that is chosen to match the lifetime. Between repetitions, the laser frequency is shifted to avoid effects of persistent spectral hole burning caused by pumping into long-lived Zeeman or hyperfine states that are split in frequency by the nuclear spins and by Earth's magnetic field (see Appendix \ref{app:PersistentSpectralHoleBurning}).

The data is fit with inverted Lorentzian curves, resulting in narrow homogeneous linewidths down to $\SI{9(3)}{\kilo\hertz}$ for the CVD sample (as shown in Fig.~\ref{fig:PowerDependence}, right inset). Comparable values are obtained for the other samples (see Appendix \ref{app:LineWidth}). The measured value is an ensemble average; it is limited by power broadening and therefore constitutes an upper bound. This can be seen from the increase of the linewidth with excitation pulse power (main panel), which follows a square-root dependence (solid lines) up to the lowest powers that still give a detectable signal. Thus, the measured upper bound to the homogeneous linewidth is among the narrowest spectral features measured in nanostructured material so far \cite{zhong_emerging_2019}, and approaches the lifetime limit of $\sim 0.5\,\si{\kilo\hertz}$ in the CVD samples.  

Still, similar to all other solid-state quantum emitters \cite{wolfowicz_quantum_2021} --- including erbium in other hosts \cite{liu_spectroscopic_2005, zhong_emerging_2019} --- we expect to observe spectral diffusion of a few $\si{\mega\hertz}$ on longer timescales, caused by the nuclear spin bath and by charge traps at the waveguide interface. Avoiding the former will be possible in isotopically purified $^{28}\text{Si}$, while the latter can be reduced via waveguides or resonators with larger dimensions \cite{merkel_coherent_2020}, by suited surface termination (e.g. by hydrogenation), and by charge-trap depletion in strong electric fields \cite{anderson_electrical_2019}.

\section{Temperature dependence} \label{sec:TemperatureDependence}

\begin{figure}[tb]
\includegraphics[width=1.0\columnwidth]{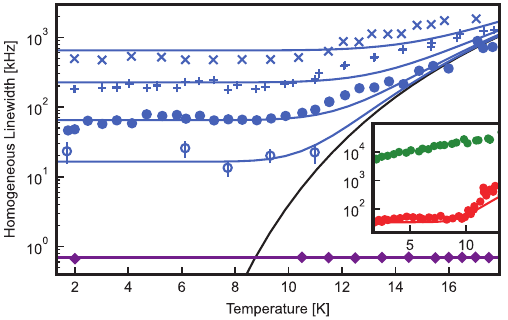}
\caption{     \label{fig:Temperature dependence}
\textbf{Temperature dependence.} The temperature of the CVD sample is changed with a resistive heater. The lifetime $\tau$ of the optically excited state, and thus the lifetime limit $1/(2 \pi\tau)$ (purple data and line) of the homogeneous linewidth, is independent of temperature, as expected for erbium dopants in the absence of non-radiative decay of the optically excited state. In contrast, the homogeneous linewidth shows an increase with temperature. For site A (blue, main panel) and several input powers (different symbols), the rise is well explained by Orbach relaxation of the spin levels (blue fit curves). From the Orbach coefficient, in the absence of power broadening (black) we expect that the lifetime-limited linewidth may be observed below $\sim 8\,\si{\kelvin}$. While site B shows similar temperature dependence (red, inset), the other bright site (green, inset) exhibits much broader lines even at $\SI{2}{\kelvin}$, owing to its smaller CF level separation.
}
\end{figure}

In addition to the fast radiative decay, we expect that the sites A and B keep their coherence when increasing the temperature beyond $\SI{2}{\kelvin}$. The reason is that both the CF separation and the Debye temperature in silicon, which determine the Orbach and Raman relaxation time, are considerably larger than in other hosts for erbium dopants (see Appendix \ref{app:OtherHostMaterials}). To investigate this, we increase the sample temperature of the CVD sample with a resistive heater and measure the lifetime and the homogeneous linewidth. The former is independent of temperature, whereas an increase is observed in the latter, as shown in Fig.~\ref{fig:Temperature dependence}. For the precipitate site (green, inset), the small separation of the CF levels entails a linewidth of $\sim\SI{5}{\mega\hertz}$ at $\SI{2}{\kelvin}$, which further increases with temperature. In contrast, we find that the measured value for the two preferred sites A (blue, main graph) and B (red, inset) is more than hundredfold smaller and constant at low temperature. It is only increased beyond the limit of our measurement setup above $\sim 10\,\si{\kelvin}$. For several settings of the optical intensity, the temperature dependence is well fit by an Orbach process, in which the exponent is determined by the independently measured crystal field splittings (see Appendix \ref{app:HomLinewidthTemperature}). Thus, only the Orbach coefficient of $\SI{0.17(6)}{\second^{-1}\kelvin^{-3}}$ (site B: $\SI{0.15(6)}{\second^{-1}\kelvin^{-3}}$) and the power-broadened linewidth are free parameters in the depicted fit curves.

In the apparent absence of Raman decoherence, the determined Orbach coefficients enable negligible decoherence of the optical transition and second-long lifetimes of the electronic spin already at $4\,\si{\kelvin}$, i.e. in a temperature range that is conveniently accessible with dry $^4\text{He}$ cryostats. This is a huge advantage with respect to other emitters in silicon that are broadened to $0.2\,\si{\giga\hertz}$ at this temperature \cite{higginbottom_optical_2022}.

\section{Discussion} \label{sec:Discussion}

In summary, we have observed exceptional optical properties of erbium dopants in silicon nanophotonic waveguides. In contrast to other emitters in silicon \cite{morse_photonic_2017, chartrand_highly_2018, durand_broad_2021, redjem_single_2020, higginbottom_optical_2022}, their weak coupling to phonons avoids non-radiative decay, phonon-sideband-emission, and thermal broadening even at temperatures around $\SI{4}{\kelvin}$. Compared to erbium in other nanostructured materials \cite{liu_spectroscopic_2005, chen_parallel_2020, craiciu_multifunctional_2021}, we observe much faster optical transitions, smaller inhomogeneous broadening, and a remarkably narrow homogeneous linewidth.

Longer waveguides, or waveguides with larger dopant concentration, should enable long-lived, chip-based quantum memories in robust, cost-effective, and multiplexed devices. In this context, our measurements set a lower bound on the ground-state electron spin lifetime of $\SI{0.02}{\milli\second}$ at $8\,\si{\kelvin}$. At lower temperature, exponentially longer lifetimes are expected. At the used concentration, dynamical decoupling can thus give access \cite{merkel_dynamical_2021} to quantum memories with long storage time. Alternatively, the large CF splitting of the observed erbium sites entails that second-long hyperfine coherence should be achievable in large magnetic fields \cite{rancic_coherence_2018}. 
While we have used silicon with natural isotope abundance, isotopic purification is possible even at wafer-scale \cite{mazzocchi_99992_2019}, and in photonic SOI \cite{liu_28silicon--insulator_2022}. This may reduce the detrimental effect of superhyperfine couplings on the coherence, and open exciting prospects for ultra-narrow optical linewidths \cite{chartrand_highly_2018} in the telecom C-band. This would help the implementation of efficient schemes for microwave-to-telecom transduction \cite{obrien_interfacing_2014} in a silicon-based platform, which is compatible and routinely used with superconducting qubits. In this context, the fraction of implanted dopants that are integrated at the respective sites will be relevant. An order-of-magnitude estimation based on comparing the detected fluorescence to the number of implanted dopants gives $\gtrsim 1\,\%$ (see Appendix \ref{app:IntegrationEfficiency}). This value may be further improved by optimizing the sample implantation and annealing procedure.

Finally, single dopants may be resolved by integration into photonic crystal \cite{chen_parallel_2020} or Fabry-Perot \cite{merkel_coherent_2020} resonators. With the low loss observed in our material, we have realized resonators with a mode volume around a cubic wavelength and a quality factor of $Q>5\cdot10^{4}$. Further improvement of the latter \cite{asano_photonic_2018} or further reduction of the former \cite{hu_experimental_2018}, both recently demonstrated in undoped silicon, may then allow for unprecedented Purcell enhancement by six orders of magnitude while preserving the unique potential of frequency-domain multiplexing \cite{chen_parallel_2020} within a well-controlled ensemble of emitters. This opens promising perspectives for the implementation of cavity-based quantum networks \cite{reiserer_cavity-based_2015, ruf_quantum_2021} and distributed quantum information processors in a scalable photonic platform.

\section{Acknowledgements}
We thank Manuel M\"uller and Stephan Gepr\"ags for assistance with the XRD measurements, and Ulrich Kentsch and Shavkat Akhmadaliev for the high-energy implantation that was carried out at ELBE at the Helmholtz-Zentrum Dresden-Rossendorf. This project received funding from the European Research Council (ERC) under the European Union's Horizon 2020 research and innovation programme (grant agreement No 757772), and from the Deutsche Forschungsgemeinschaft (DFG, German Research Foundation) under Germany's Excellence Strategy - EXC-2111 - 390814868 and via the project RE 3967/1.

\appendix

\section{Sample preparation} \label{app:SamplePreparation}

In this section, the sample preparation is described in more detail. An overview over the different sample parameters is given in Table \ref{Table_SampleOverview}.

\begin{table*} [tb] 
\begin{center}
\begin{tabular}{ c || c | c | c} 
 Sample type & CZ & CVD & FZ \\ \hline \hline
 Manufacturer / vendor & Silicon Valley Microelectronics & University Wafer and LSRL & Ultrasil \\
 DL thickness [$\si{\micro\meter}$] & 0.22 & 0.19 & 2\\  
 BOX thickness [$\si{\micro\meter}$] & 3 & 2 & 1 \\
 resistivity and dopant  [$\si{\ohm\centi\meter}$] & 10(2) (boron) & unknown (undoped) & $> 10000$ (undoped) \\
 Erbium implantation dose [$10^{12}\,\si{\centi\meter}^{-2}$]& 3.6 & 2.9 & 2 \\
 Implantation energy [MeV] & 0.07 and 0.16 and 0.35 & 0.1 and 0.35 & 1.5 and 2.5 and 4.0 \\
 Simulated peak concentration [$10^{18}\,\si{cm}^{-3}$] & 0.2 & 0.2  & 0.02 \\
 Isotopic content & $^{170}\text{Er}$ & $^{170}\text{Er}$ and $^{167}\text{Er}$ (50:50) & all isotopes \\
 Annealing temperature [K] & 800 & 800 & 800 \\
 Annealing condition & during implantation & during implantation & after implantation,  $\SI{1}{\minute}$  
\end{tabular}
\caption{Overview over the geometry and implantation conditions of the three different sample types.} \label{Table_SampleOverview}
\end{center}
\end{table*}

\subsection{Silicon substrates} \label{app:SiliconSubstrates}
\begin{figure*}[tb]
\includegraphics[width=2.\columnwidth]{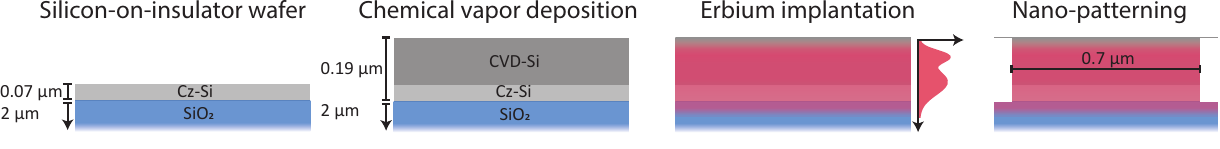}
\caption{\label{fig:SamplePreparation}
\textbf{Preparation of the CVD samples.} A commercial SOI wafer with a device layer thickness of $0.07\,\si{\micro\meter}$ is exposed to a hydrogen bake. Subsequently, pure silicon is grown by chemical vapor deposition to a total DL thickness of $0.19\,\si{\micro\meter}$. In a third step, erbium is implanted with two energies that give an approximately homogeneous dopant concentration in the device layer. Finally, established techniques (electron-beam lithography and reactive ion etching) are used to fabricate nanophotonic waveguides with a width of $0.7\,\si{\micro\meter}$. The other samples are prepared in the same way, except that different starting material is used and the CVD step is omitted.
}
\end{figure*}

For the FZ and CZ samples, we use commercial wafers from Ultrasil and Silicon Valley Microelectronics, respectively. The fabrication of the CVD samples is shown in Fig.~\ref{fig:SamplePreparation}. It starts from commercial, slightly p-doped ($R=1-20\,\text{Ohm} \cdot \text{cm}$) silicon-on-insulator wafers (University Wafer) with a $0.07\,\si{\micro\meter}$ device layer (DL), a $2\,\si{\micro\meter}$ buried oxide (BOX), and a diameter of $20\,\si{\centi\meter}$. The samples are exposed to a one-minute high-temperature hydrogen bake to evaporate the top oxide and thus prepare a clean substrate for a subsequent $\sim0.14\,\si{\micro\meter}$ crystal growth by chemical vapor deposition (Lawrence Semiconductor Research Lab, LSRL). Optical inspection revealed no misfits, and X-ray measurements confirmed the single-crystalline structure of the epitaxial layer, with a low lattice mismatch ($< 10^{-4}$) between the silicon substrate and the seed and epitaxial layers. The latter has impurity concentrations of $<5\times10^{17}\,\si{\centi\meter}^{-3}$ and $<10^{17}\,\si{\centi\meter}^{-3}$ for oxygen and carbon (measured by secondary-ion mass spectrometry), and $\lesssim 10^{14}\,\si{\centi\meter}^{-3}$ for other impurities (manufacturer specification).

\subsection{Implantation} \label{app:Implantation}
The aim of the implantation procedure is to avoid excess crystal damage and to achieve a homogeneous erbium density throughout the DL, which gives the strongest fluorescence signal at a given peak concentration. In contrast to most preceding works on Er:Si, no co-implantation of oxygen is used. The reason is that instead of a high fluorescence signal originating from Er-O clusters after off-resonant excitation, we were aiming for a high concentration of erbium dopants in a single site after resonant excitation.

The CVD samples are implanted by a commercial supplier (Ion Beam Services). To this end, they are first laser-cut to a smaller diameter of $10\,\si{\centi\meter}$ (LaserCut Processing GmbH) to fit the implantation chamber. They are then heated to $800\,\si{\kelvin}$ and implanted  under an angle of $7^\circ$ to prevent channeling. Two runs for each of the isotopes $^{167}\text{Er}$ and $^{170}\text{Er}$ are used, one with an energy of $0.1\,\text{MeV}$ and a dose of $4.5\,\times 10^{11}\,\si{\centi\meter}^{-2}$, and one with $0.35\,\text{MeV}$ and a dose of $ 10^{12}\,\si{\centi\meter}^{-2}$. Numerical simulations (SRIM.org) with a heuristic correction factor \cite{palmetshofer_range_2001} lead to an approximately homogeneous dopant concentration of $10^{17}\,\si{cm}^{-3}$, as shown in the Fig.~\ref{fig:Waveguides}a and b.

The CZ samples are implanted by the same supplier and process, except that the wafers are diced before implantation and only $^{170}\text{Er}$ is used. Furthermore, three runs are performed to improve the homogeneity of the concentration while keeping the same target peak concentration.

The FZ samples are much thicker and thus need higher implantation energies, available from the Tandetron accelerator at ELBE at the Helmholtz-Zentrum Dresden-Rossendorf. Again, three implantation runs with an angle of $7^\circ$ and different energies were chosen to give an approximately homogeneous concentration in the DL. Isotope selection and heating during the implantation procedure was not possible. To heal the implantation damage, the samples were therefore annealed afterwards in a home-built rapid thermal annealing oven. The temperature is ramped up to $\SI{800}{\kelvin}$ within three minutes, kept there for one minute, before cooling down with an initial ramp rate of $\SI{7}{\kelvin / \second}$. Since at the used temperature, the mobility of erbium in silicon is very small \cite{kenyon_erbium_2005}, we do not expect that the annealing significantly changes the spatial erbium distribution in the samples. However, the site occupation will depend on the exact annealing procedure, and may thus be optimized in a future systematic study.

\subsection{Waveguide fabrication} \label{app:WvgFabrication}
The photonic nanostructures are fabricated by electron beam lithography using ZEP resist on a nB5 machine (Nanobeam Ltd.). The pattern is transferred via reactive ion etching (Oxford Plasmalab 80) using a mixture of $\text{SF}_6$, $\text{O}_2$ and Ar etch gases at cryogenic temperature.

On the CZ and CVD samples, the fabricated structures are ridge waveguides with a width of $\SI{700}{\nano\meter}$, which support three modes, one with TM and two with TE polarization. The waveguides are terminated with a broadband photonic crystal reflector for TE polarized light in order to monitor the incoupling and collect light emitted into both waveguide directions. The reflectors consist of 30 holes with a separation of $\SI{330}{\nano\meter}$ and a design diameter of $\SI{150}{\nano\meter}$ to $\SI{180}{\nano\meter}$, which reflects the different device layer thickness of the CVD and CZ samples.

To maximize the signal and thus be sensitive to small erbium dopant ensembles, we need a highly efficient and broadband off-chip coupling, which is best achieved in a side-coupling geometry \cite{vivien_handbook_2013}. To this end, we terminate the waveguide in a $\SI{30}{\micro\meter}$ long inverted taper with a tip diameter of below $\SI{200}{\nano\meter}$ and couple to it via a tapered single-mode fiber, fabricated by etching in hydrofluoric acid \cite{nikbakht_fabrication_2015}. Using the setup described in the next section, the fiber is aligned with the sample under optical inspection. While this allows for a coupling efficiency approaching unity for underetched, free-standing structures \cite{sipahigil_integrated_2016}, for structures that reside on a $\mathrm{SiO}_2$ BOX layer we simulated a coupling of up to $0.3$ for an optimized geometry, and typically observe around $0.1$ in the experiments.

The FZ waveguides have a larger cross-section and thus favor a different coupling scheme. They are thus terminated by polished edges, to which a single-mode fiber (SMF-28) is glued using UV-curing, low-outgassing epoxy (Norland NOA 88). The mode-mismatch between the fiber leads to a moderate coupling efficiency, estimated around $4\,\%$, but also ensures that the coupling is insensitive to displacements caused by thermal contraction upon cooldown. 

\section{Experimental setup}  \label{app:ExpSetup}

\begin{figure*}[tb]
\includegraphics[width=2\columnwidth]{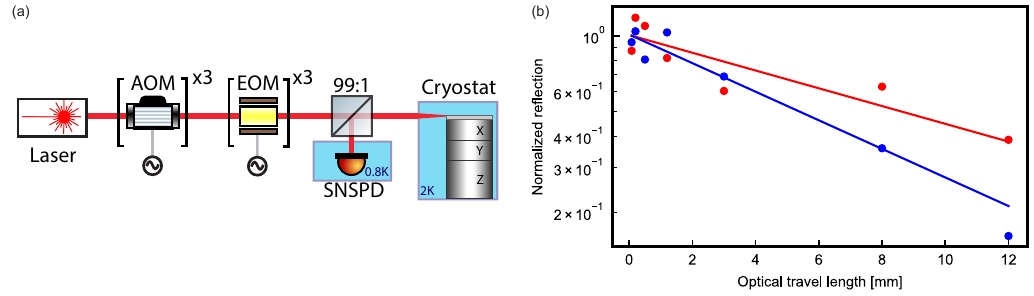}
\caption{     \label{fig:Setup_Loss}
\textbf{a) Experimental setup.} Acousto-optical modulators (AOM) are used to generate pulses of controlled duration, amplitude, and frequency shift from a continuous-wave laser. Electro-optical modulators (EOM), driven by an arbitrary-waveform-generator (AWG), are used to generate optical sidebands. A fiber-based non-polarizing beam-splitter with a $99:1$ (in some measurements $95:5$) split ratio is used to irradiate the excitation light, and guide the fluorescence signal to a superconducting nanowire single-photon detector (SNSPD). The sample is kept in a closed-cycle cryostat on an XYZ nanopositioning stage.
\textbf{b) Waveguide propagation loss on the CVD samples.} The reflected signal is measured for waveguides of different length. An exponential fit then allows for extracting the propagation loss for doped (blue) and undoped (red) CVD samples. 
}
\end{figure*}

A sketch of the experimental setup is shown in Fig.~\ref{fig:CrystalFieldLevels}. Fig.~\ref{fig:Setup_Loss}a gives an enlarged view. The samples are mounted on a nanopositioning platform (Attocube) in a closed cycle cryostat (Attocube Attodry 2100) and cooled in helium buffer gas to temperatures below $2\,\si{\kelvin}$. The erbium dopants are excited with laser pulses generated from a continuous-wave source (Toptica CTL) via three fiber-coupled acousto-optical modulators (Gooch and Housego). The laser frequency is determined to an absolute precision of $20\,\si{\pico\meter}$ using a calibrated spectrum analyzer (Bristol Instruments 771 series). For sweeping the laser frequency over larger ranges, the angle of the grating in the laser cavity is scanned. A small added modulation avoids a reduction of the fluorescence level which we attribute to persistent spectral hole burning. When not swept, the laser is stabilized to a precision of $\sim3\,\si{\kilo\hertz}$ using a frequency comb (Menlo Systems FC-1500-250-ULN). Then, the phase of the light field is modulated by electro-optcial modulators (Ixblue), and precise frequency offsets are added by the acousto-optical modulators using a home-built radio-frequency synthesizer or an arbitrary waveform generator (Zurich Instruments HDAWG).

The fluorescence signal is detected with superconducting-nanowire single-photon detection systems with an efficiency up to $80\,\%$ at less than $100\,\text{Hz}$ dark count rate (Photon Spot and ID Quantique). To determine the wavelength of the emission, we use an electrically tunable fiber-based filter (WL photonics WLTF-NE-S-1550) with a Gaussian transmission of $\SI{0.11}{\nano\meter}$ FWHM, which we calibrate using the wavemeter.

Using fiber-based polarization controllers, the polarization at the CVD and CZ samples is aligned such that the reflected signal is maximized. From simulations of the mirror, we expect that this corresponds to alignment with the fundamental TE mode of the waveguide \cite{vivien_handbook_2013}. For the FZ samples, the polarization is in a random state.

\section{Waveguide loss}  \label{app:WvgLoss}
To determine the propagation loss constant $\alpha$ in the CVD waveguides, we fabricate many structures with different waveguide lengths on a single chip and measure the fraction of the intensity $I$ that is reflected at room temperature: 
\begin{equation}
    I_\text{out}/I_\text{in} = R \eta^2 \text{e}^{-2\alpha L}. 
\end{equation}

Here, $\eta$ is the chip-to-fiber coupling efficiency, $R$ the reflectivity of the photonic crystal mirror and $L$ the waveguide length. $\eta$ and $R$ may fluctuate because of fabrication imperfections, and have a maximal value $\eta_\text{max}$ and $R_\text{max}$ only for perfect structures. Thus, for each value of $L$ we measure the reflection of several waveguides averaged over the accessible frequency range. For analysis of the loss, we then discard all measurements except for the largest values that are obtained when $\eta \simeq \eta_\text{max}$ and $R \simeq R_\text{max}$. We then fit the remaining data set to an exponential decay, as shown in Fig.~\ref{fig:Setup_Loss}b. We find $\alpha=(5.7 \pm 1.2)\,\text{dB/cm}$ for the doped CVD samples. For comparison, using the same recipe we also fabricate waveguides on undoped CZ samples, which leads to a slightly lower loss of $\alpha=(3.5 \pm 1.2)\,\text{dB/cm}$. As a second loss characterization technique, we fabricate ring-resonators on the doped CVD samples. We observe resonant modes with a $Q$-factor of $\sim 10^5$ for a round-trip length of $\SI{70}{\micro\meter}$. From that, we can estimate an upper limit of $\alpha<8(1)\,\text{dB/cm}$ \cite{bogaerts_silicon_2012}, within errors in agreement with the previous result.

The slight increase of the propagation loss in the implanted samples may be caused by residual crystal damage from the implantation process, and may thus be further reduced in samples with optimized implantation conditions.

\section{Crystal field levels}  \label{app:CrystalFieldLevels}
\begin{figure*}[tb]
\includegraphics[width=2 \columnwidth]{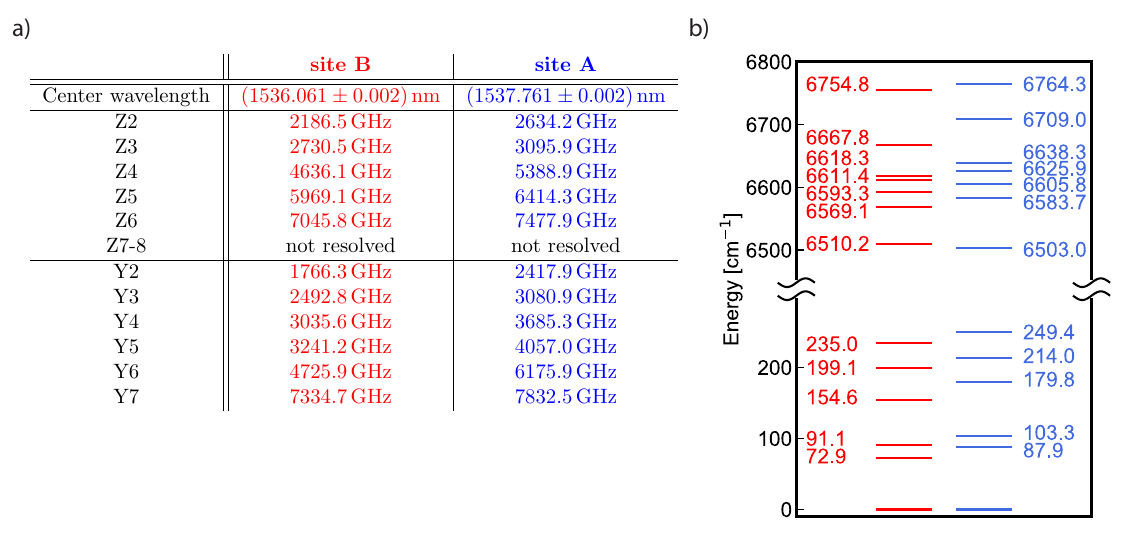}
\caption{\label{fig:LevelScheme} \textbf{Energy of the crystal field levels of sites A (blue) and B (red).} a) Frequency difference of the CF levels with respect to the ground state of the respective manifold. b) Level scheme with the corresponding wavenumbers. The excited state includes all CF levels of the $^4I_{13/2}$ state manifold, whereas in the $^4I_{15/2}$ ground state manifold, two levels have not been detected because the used filter only tunes to $\SI{1600}{\nano\meter}$.
}
\end{figure*}

When rare-earth dopants are integrated into host crystals, the degeneracy of their 4f levels is lifted by the crystal field (CF) \cite{liu_spectroscopic_2005}. The transitions between these levels can be resolved by spectroscopy, as described in Section \ref{sec:CrystalFieldMeasurements}. The extracted values of the CF separation are summarized in Fig.~\ref{fig:LevelScheme}.

\section{Comparison between CVD, FZ and CZ samples.}
Except in the instances shown in Fig.~\ref{fig:InhomLine_Purcell_Lifetime}, the properties of the erbium dopants do not show large differences between the different sample types. Still, in the following sections, the measurements on the other samples are included to allow for a detailed comparison between the different sample types.

\subsection{Fluorescence spectrum}  \label{app:FluorescenceSpectrum}

\begin{figure*}[tb]
\includegraphics[width=2 \columnwidth]{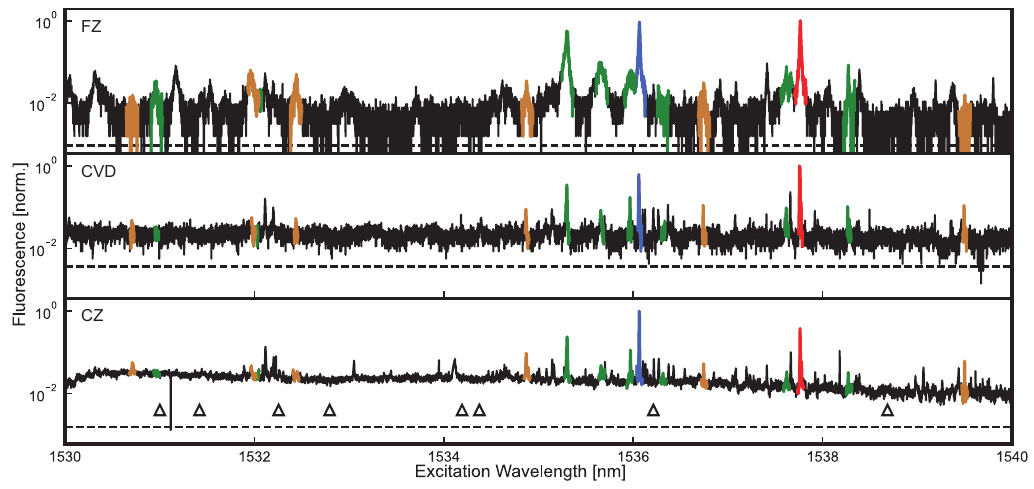}
\caption{\label{fig:CompareSpectrum} \textbf{Comparison of the resonant excitation spectrum in the different sample types}. Compared to the CVD and CZ samples, the FZ chips exhibit a smaller number of resonant peaks, which is attributed to the higher purity of the starting material. The triangles at the bottom of the third panels indicate different sites that were observed in our earlier work \cite{weiss_erbium_2021}, which used different implantation conditions.
}
\end{figure*}

Fig.~\ref{fig:CompareSpectrum} shows the emission as a function of the resonant excitation wavelength between $\SI{1530}{\nano\meter}$ and $\SI{1540}{\nano\meter}$ (i.e. in the range where the transition between the lowest CF levels is expected) for all three sample types. The same measurement technique as in Fig.~\ref{fig:CrystalFieldLevels} has been applied. To avoid an offset from the reflected excitation laser, we use excitation pulses of $\SI{5}{\micro\second}$ (in the CZ and FZ samples) or $\SI{0.05}{\micro\second}$ (CVD) duration. After this, the fluorescence is compared in a time interval of $\SI{0.25}{\milli\second}$. From measurements on the same sample, we find that the pulse duration does not have a significant influence on the obtained spectrum. On each wavelength setting, 500 (CVD), 2000 (CZ) or 2500 (FZ) averages are acquired. To avoid effects of persistent spectral hole-burning (discussed below), the laser frequency is modulated on a ms-timescale around its setpoint via a piezo that changes the angle of a built-in grating. Thus, the excitation pulse spectrum is broadened to approximately $\SI{0.05}{\giga\hertz}$, which is smaller than the inhomogeneous broadening, but large enough to address both Zeeman states (that are split by earths' magnetic field), and also slightly larger than the step size of the sweep (which ensures that also ultra-narrow features would not be missed). The normalized spectra are plotted on a logarithmic scale. In all samples, we observe a broadband background discussed below. This background exceeds the dark count level (dashed lines) in all samples. Its relative amplitude is significantly smaller in the FZ samples, albeit the peaks to which the spectrum is normalized are broader because of the higher implantation energies. The decreased background may be attributed to the changed waveguide geometry, or to the higher purity of the FZ material, or the reduced erbium density.

\subsection{Inhomogeneous and homogeneous linewidth}  \label{app:LineWidth}

\begin{figure}[t]
\includegraphics[width=1\columnwidth]{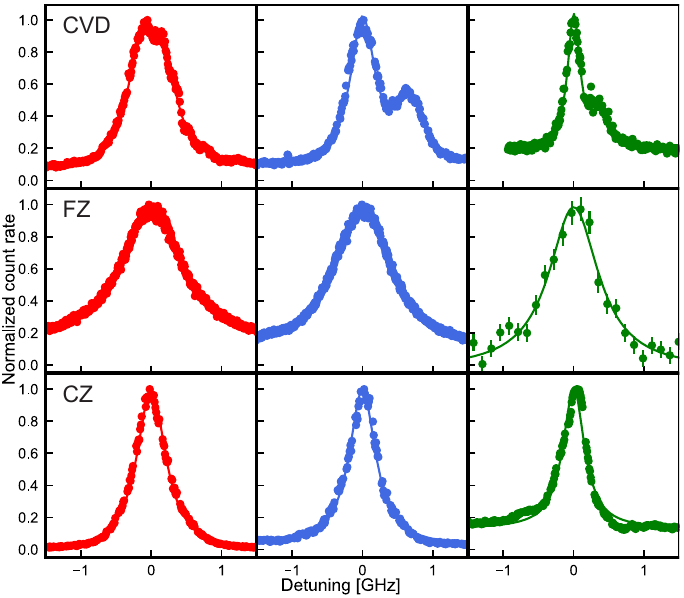}
\caption{\label{fig:CompareInhomogeneousLines} \textbf{Inhomogeneous linewidth comparison}. The narrowest lines are obtained in the CZ samples. The FZ samples show an increased linewidth that is attributed to an increased crystal damage caused by the higher implantation energy. The CVD samples show a double-peak structure, which we attribute to the two implanted isotopes.
}
\end{figure}

In Fig.~\ref{fig:CompareInhomogeneousLines}, the spectrally filtered emission is shown as a function of the detuning of the resonant excitation laser for all three sites. The measurement uses the same technique that was used in Fig.~\ref{fig:InhomLine_Purcell_Lifetime}a, except for the measurement of site P on the FZ sample, which was measured without spectral filtering because of the lower signal obtained for this peak in that sample.

In the CVD samples, a double-peak structure is observed. Fitting a double-Lorentzian with equal FWHM of both peaks gives an inhomogeneous linewidth of $0.40(1)\,\si{\giga\hertz}$ (site A), $0.44(1)\,\si{\giga\hertz}$, and $0.23(1)\,\si{\giga\hertz}$ (P), with a separation of $0.65(1)\,\si{\giga\hertz}$ (A), $0.36(1)\,\si{\giga\hertz}$ (B) and $0.32(1)\,\si{\giga\hertz}$ (P). They two peaks may originate from an inhomogeneous strain in the CVD-grown device layer, or from the two isotopes $^{167}\text{Er}$ and $^{170}\text{Er}$ that have been implanted with equal dose in these samples.

\begin{figure}[t]
\includegraphics[scale=1]{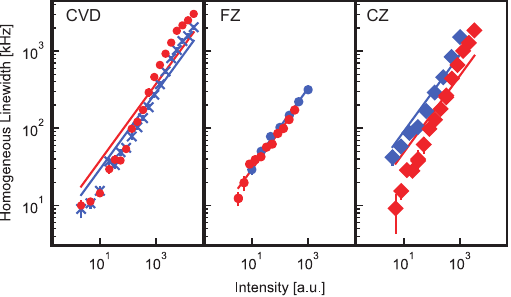}
\caption{\label{fig:CompareHomogeneousLines} \textbf{Homogeneous linewidth comparison}. In all samples, the homogeneous linewidth is limited by power broadening, even at the lowest powers that give a sufficient signal.
}
\end{figure}

Fig.~\ref{fig:CompareHomogeneousLines} shows the homogeneous linewidth as a function of the excitation laser power for all three samples. 

\section{Properties of the precipitate site "P" and the optically active site "O"}  \label{app:PandOSites}

As discussed in the Section ~\ref{sec:SamplesAndExperimentalSetup}, we not only observe fluorescence of the main sites A and B, but also smaller signals originating from several other sites. The fraction of erbium dopants in these sites may be further reduced by optimizing the implantation procedure. Still, in the following the properties of two most relevant other sites are investigated.

\begin{figure*}[tb]
\includegraphics[width=2 \columnwidth]{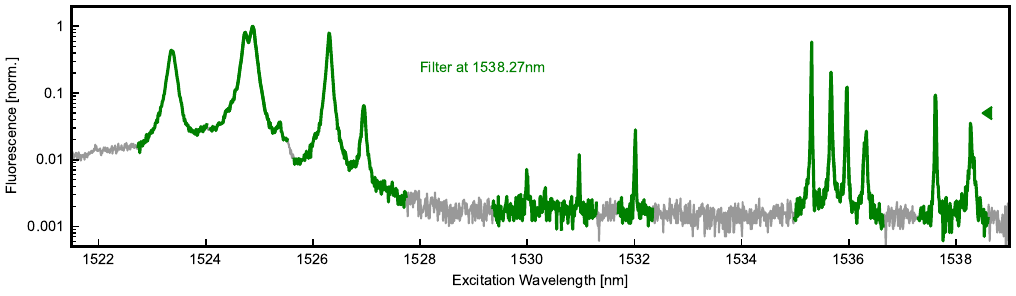}
\caption{\label{fig:PrecipitateSite}
\textbf{Excited state CF levels of site "P".} Site "P" is attributed to erbium precipitates as it shows 14 CF levels in the $I_{13/2}$ manifold --- twice the number expected for a single erbium site in the absence of a magnetic field. As the smaller CF splitting of this site cannot be resolved well in Fig.~\ref{fig:CrystalFieldLevels}b, this figure shows the data over a smaller wavelength spread. While we can resolve the $I_{13/2}$ levels well by scanning the narrow-linewidth laser, not all $I_{15/2}$ levels are resolved because of the finite $\SI{0.11}{\nano\meter}$ bandwidth of the available wavelength filter.
}
\end{figure*}

The first one, site "P", is already discussed in Section~\ref{sec:CrystalFieldMeasurements} and Fig. \ref{fig:CrystalFieldLevels} (green). Fig.~\ref{fig:PrecipitateSite} shows a zoom into its excited state CF spectrum that allows for a better resolution of the lines. The large number of CF levels indicates that dopants in site "P" occupy a pair- or precipitate site.

\begin{figure*}[tb]
\includegraphics[width=2 \columnwidth]{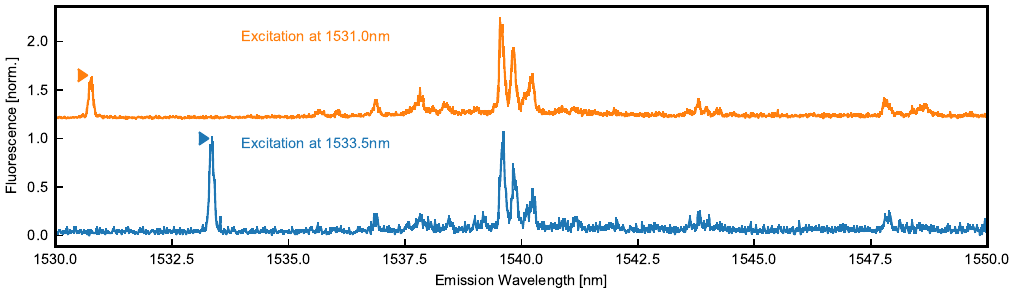}
\caption{\label{fig:OpticallyActiveSite}
\textbf{Fluorescence upon off-resonant excitation in the CVD sample.} The samples are excited at two different wavelengths (triangles), $\SI{1531.0}{\nano\meter}$ (orange, with added offset) and $\SI{1533.5}{\nano\meter}$ (blue), which do not coincide with one of the resonances in Fig.~\ref{fig:CrystalFieldLevels}a. The filtered fluorescence signal recorded after the laser pulses shows a peak at the excitation frequency, and in addition some off-resonant fluorescence at several discrete wavelengths, which we attribute to charge relaxation via an optically active site \cite{vinh_photonic_2009}.
}
\end{figure*}

In addition, we observe one (or several) sites that we call "O" as they are "optically active", which means that their fluorescence can be excited by electrons in the conduction band \cite{vinh_photonic_2009}. This entails that we observe emission of these sites independent of the excitation wavelength, which leads to a background observed in the unfiltered fluorescence shown in Fig.~\ref{fig:CompareSpectrum}. While the excitation of this background is largely independent of the laser wavelength, the emission is narrow-band, given by the CF level separation of the sites. This can be seen in Fig.~\ref{fig:OpticallyActiveSite}, where we show the emission spectrum obtained for two excitation wavelengths that are off-resonant from any of the observed resonances. 

In contrast to earlier measurements on implanted samples that revealed many sites with large inhomogeneous broadening \cite{przybylinska_optically_1996, kenyon_erbium_2005}, because of our higher silicon purity, lower dose and lower annealing temperature we also observe only few and comparably narrow lines from site(s) O, limited by the resolution of the wavelength filter. The fluorescence peaks of O do not coincide with the emission of sites A, B, and P. Centering the wavelength filter on the largest peak around $\SI{1539}{\nano\meter}$, we find that site(s) O also exhibits a different lifetime of $\sim\SI{0.7}{\milli\second}$ in the CVD sample.

In addition to the mentioned off-resonant fluorescence of site(s) O, we also observe a weak resonant fluorescence background when exciting the system between $\sim 1520$ and $1540\,\si{\nano\meter}$. Its amplitude depends on the wavelength and is comparable to the fluorescence of site O, as can be seen for two wavelengths in Fig.~\ref{fig:OpticallyActiveSite} (triangles). Its lifetime of $2.47(6)\,\si{\milli\second}$ is similar to earlier observations \cite{kenyon_erbium_2005}, and matches the expectation for a predominantly-magnetic dipole decay in a non-magnetic medium \cite{dodson_magnetic_2012}. 

\section{Modification of the lifetime by the photonic density of states}
\label{app:LDOS}

\begin{figure*}[tb]
\includegraphics[width=2\columnwidth]{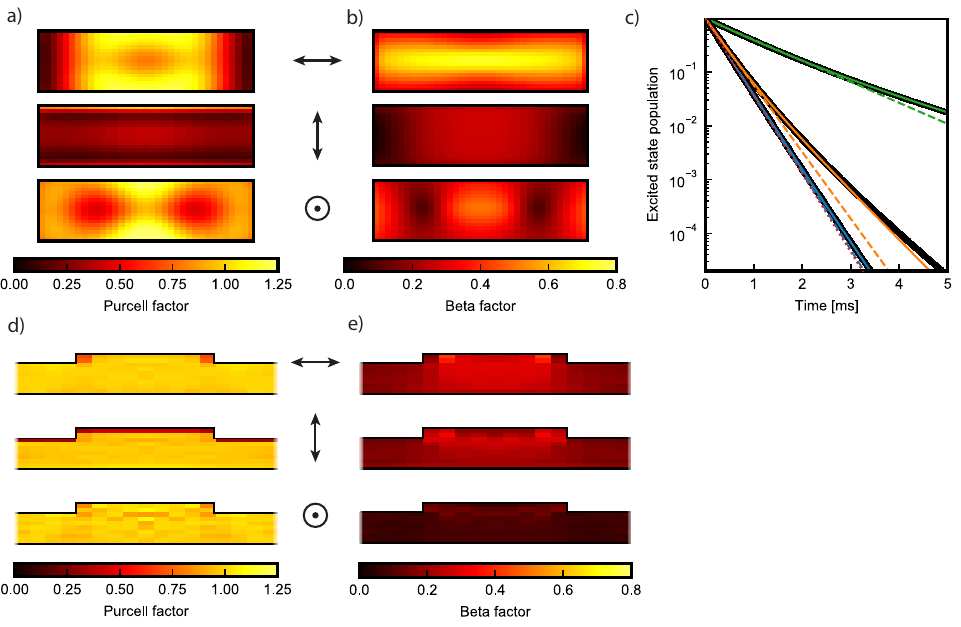}
\caption{\label{fig:Purcell_Enhancement}
\textbf{Modification of the electric-dipole emission in the different samples.}  Finite-difference-time-domain (FDTD) calculations are used to simulate the photonic density of states depending on the position in the waveguide (cross-sectional images) and the orientation of the electric dipole (black symbols). From this, the emission rate of erbium dopants is determined. In addition, the probability $\beta$ to emit into the guided mode is calculated. The dimensions of the waveguides differ between samples, as shown in Fig.~\ref{fig:Waveguides}. \textbf{a) and b): Ridge waveguides used on the CVD and CZ samples.} The decay rate can be increased (Purcell enhancement $F_P>1$) or reduced ($F_P<1$) depending on the position. $\beta$ approaches unity for erbium dopants at the waveguide center with a horizontal dipole orientation (top), but is much smaller for the other dipole orientations. \textbf{c) Expected temporal decay of the fluorescence in the ridge waveguides.} A Monte-Carlo-Simulation is performed, in which dopants are randomly positioned in the waveguide according to the implantation profile. Then, their dipole decay is averaged, weighted by the $\beta$-factor. The resulting data for different dipole orientations (black) deviate from a single exponential fit (dashed). A bi-exponential decay (solid) gives better agreement. \textbf{d) and e): Rib waveguides used on the FZ samples.}. The Purcell factor is $F_P=1$, almost independent of the position in the waveguide. Thus, one expects a decay with a time constant that is independent of the emitter orientation, and that matches the lifetime of the dopants in bulk crystals. The average $\beta$-factor is much lower than in the ridge waveguide geometry, and mostly light of transversal polarization (top and center panel) is emitted into the guided mode.
}
\end{figure*}

As discussed in Section ~\ref{sec:OpticalLifetime}, the optical emission of erbium dopants in silicon waveguides depends on the local density of states (LDOS). Albeit an exact quantitative modeling is hindered by the unknown transition dipole moments of erbium in silicon, some qualitative understanding can be gained from the simulations provided in this section. To this end, finite-difference time-domain simulations are performed with the free and open-source software MEEP (https://meep.readthedocs.io/). The results are shown in Fig.~\ref{fig:Purcell_Enhancement}, where we determine the lifetime modification (i.e. the Purcell factor $F_P$) and the emission probability $\beta$ into any of the guided modes, depending on the orientation of the radiative dipole and its position in the waveguide \cite{lodahl_interfacing_2015}. Earlier studies \cite{stepanov_quantum_2015, lodahl_interfacing_2015} have shown that the Purcell factor is not only determined by the enhancement of the radiation into the guided mode, but also by the suppression of emission into perpendicular directions. Therefore, the spatial dependence of $F_P$ in nanowaveguides can differ significantly from that of the electric field profile of the guided modes.

The CVD and CZ samples, panel a, feature waveguides of subwavelength dimension. For an electric dipole oriented parallel to the waveguide width (top and Fig.~\ref{fig:Waveguides}d), or along the propagation direction (bottom), the emission is enhanced up to 1.2-fold, depending on the emitter position. In contrast, it is suppressed to 0.3 or less when the dipole is oriented along y.

In addition to the Purcell enhancement, we also determine $\beta$, the probability that a photon is emitted into a guided mode rather than into free space. We find that in the CVD and CZ samples, panel b, photons from a dipole oriented parallel to the waveguide width (top) have a much higher probability to be emitted into one of the three guided modes of our waveguide \cite{vivien_handbook_2013}, and will thus contribute more to the signal in our measurements. However, also the dipole orientation along the waveguide will give a considerable contribution, as it has values of $F_P$ and $\beta$ that are comparable to the other orientations. This can be understood by considering that the electromagnetic field of the guided modes in a subwavelength-scale waveguide has a significant component in the longitudinal direction \cite{stepanov_quantum_2015}.

The situation is very different in the FZ samples, in which the guided mode is only weakly confined. This has two effects \cite{lodahl_interfacing_2015}: first, one would not expect a significant emission enhancement in such structures. Second, the decay into non-guided radiative modes should also be largely unaffected by the waveguide. Thus, as can be seen in Fig.~\ref{fig:Purcell_Enhancement}d, the LDOS in the FZ samples is largely independent of the emitter position and dipole orientation, and identical to that in a bulk crystal. In particular, no Purcell enhancement is expected for erbium in these samples. Considering $\beta$, panel e, one finds that most of the coupled light is emitted by dipoles with a transversal electric field polarization (top and center), in agreement with the intuitive understanding from the guided mode profile.

To model how the spatial distribution of the implanted erbium emitters will change the measured lifetime, we perform Monte-Carlo simulations. We randomly pick a dopant position according to the simulated implantation profile, and randomly assign an optical dipole orientation along one of the three principal axes shown in Fig.~\ref{fig:Purcell_Enhancement}a and b. We then determine the dipolar decay at this position. We restrict our analysis to electric-dipole transitions, which we expect to dominate the emission in the fast-decaying erbium sites A and B \cite{liu_spectroscopic_2005}. However, we note that also magnetic dipoles may contribute to the measured signal; they exhibit a similar Purcell enhancement as electrical dipoles and thus give qualitatively the same results (not shown).

When averaging the decay of a large, random erbium ensemble for the waveguide geometry in the CVD samples, weighted by the $\beta$-factor, we obtain the three curves in Fig.~\ref{fig:Purcell_Enhancement}c. The decay is well-fit by bi-exponential curves, whose fast component has a slower decay time than that in bulk (red dotted line) for all dipole orientations. This qualitatively matches the slower biexponential decay observed in the CVD and CZ waveguides as compared to the faster decay in the FZ samples. To obtain quantitative agreement, however, a better control of the dopant profile, as well as information about the orientation, shape and relative strength of the magnetic and electric dipole moments would be necessary. In addition, the different modes of the waveguide may have a different chip-to-fiber coupling efficiency, which will affect both the excitation and detection probability. 

The fact that the geometry of the waveguides has a strong effect on the lifetime indicates that the fast decay of the newly observed sites is predominantly radiative. Still, from the measurements presented so far, also an effect of the different sample preparation cannot be ruled out with certainty. Therefore, another series of measurements is performed, in which waveguides of the same thickness, but different widths are fabricated on a CZ sample. Also in this case, the fast decay constant of the biexponential fit in the lifetime measurements changes on site A from $\SI{260.5(1)}{\micro\second}$ for a waveguide of $\SI{0.58}{\micro\meter}$ design width to $\SI{209.0(1)}{\micro\second}$ for a $\SI{0.95}{\micro\meter}$ wide waveguide. On site B, the same dependence is observed: $\SI{338.1(2)}{\micro\second}$ on the narrow and $\SI{266.8(1)}{\micro\second}$ on the wider waveguide. This change by $\sim 20\,\%$ between geometries in the same sample unambiguously proves that the observed lifetime modification is not caused by a different crystal and implantation procedure, but by the variation of the radiative decay rate due to the changed local density of states.

\section{Homogeneous linewidth measurements}  \label{app:HomLinewidth}

A wide-spread technique to characterize the homogeneous linewidth in rare-earth dopant ensembles is photon echo \cite{liu_spectroscopic_2005}. In this technique, an optical $\pi/2$ pulse prepares an optical coherence that quickly dephases because of the inhomogeneous broadening of the ensemble. An optical $\pi$ pulse then causes a rephasing of the spins, leading to the emission of an echo signal. The decay of this echo as a function of the delay between the pulses then allows for determining the homogeneous linewidth.

\begin{figure}[b]
    \centering
    \includegraphics[width=1\columnwidth]{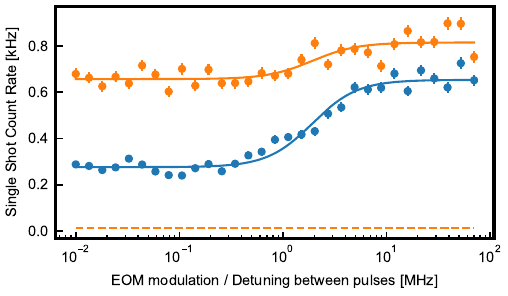}
    \caption{\label{fig:HomLW_aom_vs_rogge} \textbf{Comparison of two different techniques to measure the homogeneous linewidth.} At the same pulse parameters, the technique introduced in \cite{weiss_erbium_2021} (blue data and Lorentzian fit with FWHM $\SI{3.8(1)}{\mega\hertz}$) and used in Section~\ref{sec:HomogeneousBroadening} gives a significantly better signal-to-noise ratio than the pump-probe scheme introduced in \cite{berkman_sub-megahertz_2021} (orange data and fit with FWHM $\SI{3.8(4)}{\mega\hertz}$). The y axis shows the detected photons per repetition, divided by the detection time window.}
\end{figure}

In our setup, this measurement technique is hindered by several complications: First, photon echo is a collective effect. Thus, it does not work well in samples with only a few emitters. Second, the bandgap of silicon is much lower than that of previously studied erbium hosts. The irradiation of strong laser pulses can thus lead to the generation of free carriers and excitons via two-photon absorption. These may in turn change the status of charge traps which --- via the Stark effect --- alter the emission frequency of embedded dopants. The resulting instantaneous spectral diffusion may preclude the observation of an echo signal. The third complication of the photon echo technique is the superhyperfine interaction with the $^{29}\text{Si}$ nuclei in the host, which will lead to a fast initial collapse of an echo signal when using strong, broadband excitation pulses \cite{car_superhyperfine_2020}. In principle, this effect can be eliminated in high magnetic fields \cite{bottger_effects_2009}, by using hyperfine states of $^{167}\text{Er}$ \cite{ortu_simultaneous_2018}, and potentially in isotopically purified silicon samples. Thus, it is common to define the slow decay as the homogeneous linewidth. Still, detecting a photon-echo in the weakly-doped samples studied at zero field in this work is hindered by the fact that most of the signal would decay very rapidly.

The mentioned limitations are avoided when using weak and narrowband excitation pulses. We therefore characterize the homogeneous linewidth via transient spectral hole burning \cite{szabo_observation_1975, volker_hole-burning_1989} instead of photon echo. As can be seen in Fig.~\ref{fig:HomLW_aom_vs_rogge}, we first compare two techniques: The first is a pump-probe scheme introduced and explained in \cite{berkman_sub-megahertz_2021}, in which two pulses of $\SI{12.5}{\micro\second}$ duration are irradiated one after the other, and subsequently the fluorescence is measured for $\SI{1.5}{\milli\second}$ as a function of the detuning between the pulses. When simultaneously irradiating several far-separated laser fields, the signal can be improved while obtaining the same linewidth (not shown). In the depicted measurement, we therefore use three lines with $\SI{80}{\mega\hertz}$ separation, generated by an EOM.

This also allows for a direct comparison with the second scheme, which is described and used in Section~\ref{sec:HomogeneousBroadening}. It is also a pulsed technique, in which the fluorescence is detected after the pulses have been switched off. Using the same total excitation pulse duration, power and detection interval as in the first measurement, as expected the method gives the same homogeneous linewidth (blue vs orange curve in Fig.~\ref{fig:HomLW_aom_vs_rogge}), which is given by half the width of the transient holes \cite{volker_hole-burning_1989}. As a better contrast and thus a higher sensitivity is obtained by the latter technique, it is used throughout the manuscript.

\begin{figure*}[tb]
    \centering
    \includegraphics[width=2\columnwidth]{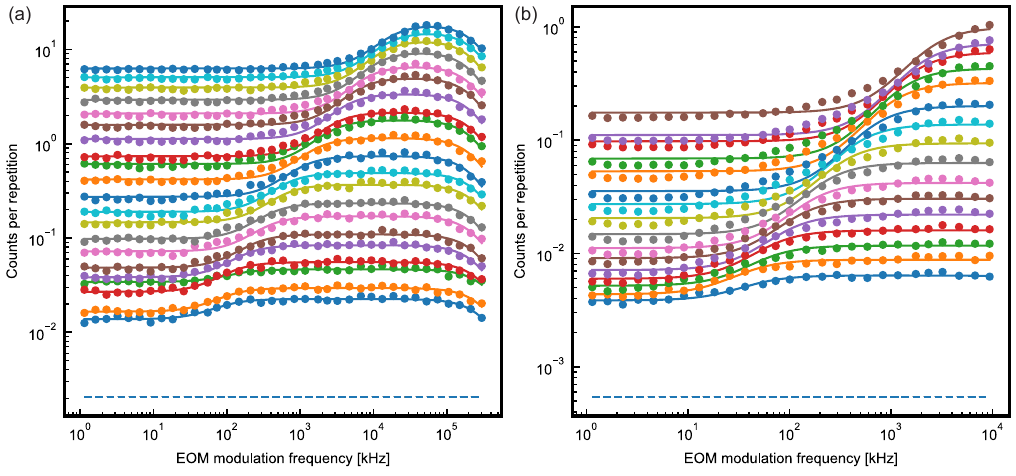}
    \caption{\label{fig:HomLW_AllData} 
    \textbf{Measurement of the homogeneous linewidth using one EOM (a) or three concatenated EOMs (b) to generate 3 or 27 lines.} The different randomly chosen colors indicate measurements at different power. The larger the power, the larger the power-broadened homogeneous linewidth, the larger the number of excited dopants, the larger the fluorescence. At the same power-broadened homogeneous linewidth, a better signal-to-noise ratio is obtained when using three EOMs instead of only one.}
\end{figure*}

In the following, the measurement technique is described and analyzed in more detail. It is based on the observation that the fluorescence signal $S$ increases nonlinearly with the laser intensity $I$ applied at a single frequency, $S \propto \sqrt{I}$, because of saturation. In contrast, a linear increase is observed when adding a second, or several, fields with a separation that is much larger than the homogeneous linewidth. A similar method has been introduced in \cite{szabo_observation_1975}, where a small amplitude modulation generated by an EOM was combined with heterodyne detection. In \cite{weiss_erbium_2021}, we adapted the technique to give a good signal-to-noise ratio in fluorescence originating from waveguide-coupled dopants, by using single-photon counters and a phase-modulating EOM to generate several fields with defined frequency separation from a single laser. The driving AC voltage is adjusted such that the carrier and first sidebands have approximately equal intensity. The second and higher-order sidebands then have much lower intensity and are therefore neglected. In this scenario, one expects an approximately threefold increase of the fluorescence (as compared to a single field) when the detuning of the laser fields is larger than the ensemble homogeneous linewidth. If, however, the detuning is small (or zero), the fluorescence will only increase by a factor of $\sqrt{3}$. 

In the experiment, at high powers and low temperature one can observe a slightly larger increase than the ideally-expected factor of $\simeq 1.73$ as a result of persistent spectral holeburning, as detailed in the next section. In addition, one observes a smaller increase under three conditions: First, if the laser fields do not have perfectly equal intensity. Second, when large modulation frequencies (or many more than three lines) are used such that the reduced dopant spectral density at the side of the inhomogeneous broadening becomes relevant. Finally, a deviation is observed at low intensities, when the resonant fluorescence signal approaches the background that originates from detector dark counts and off-resonant fluorescence, and when the dopant spectral density is smaller than one emitter per homogeneous linewidth.

These effects can be seen in Fig.~\ref{fig:HomLW_AllData}, where the raw data of the measurements on the CVD sample and the corresponding fits are shown, using either one EOM to approximately generate three laser lines (panel a), or three EOMS to generate 27 lines (panel b, c.f. Fig.~\ref{fig:PowerDependence}). Using one EOM, the expected signal increase of $\sqrt{3}$ is observed at intermediate power. Using three EOMs, the contrast is improved, but the relative improvement is smaller than the ideally expected factor of $\sqrt{27}/\sqrt{3}=3$. This is explained by the higher-order sidebands of one EOM, which can interfere with the first-order sidebands of another. This leads to unequal intensities of the 27 applied laser fields. Still, the improvement is large enough to increase the measurement range to lower homogeneous linewidths. At the lowest powers, the modulation depth decreases significantly because of the finite dopant spectral density, even before the effect of dark counts (dashed lines) becomes significant. Thus, accurately determining the homogeneous linewidth will require samples with increased dopant spectral density.


\section{Signatures of persistent spectral hole burning} \label{app:PersistentSpectralHoleBurning}

The large crystal field splitting of the newly-observed sites leads to a small spin-lattice relaxation via the Orbach process. At a temperature of $\SI{2}{\kelvin}$, this entails a long lifetime of the spin states, which are split in Earths' magnetic field. In case the splitting is different in the ground and excited state, as it is typically observed in rare-earth dopants in low-symmetry sites, the transition frequency of the dopants will change when they are pumped from one spin state to the other. This effect is known as persistent spectral hole burning \cite{liu_spectroscopic_2005}. It can strongly reduce the fluorescence signal at a given frequency. Therefore, to avoid a loss in signal, the frequency of the excitation pulses is changed between repetitions to one out of 20 random frequency shifts in the range of up to $\SI{100}{\mega\hertz}$. 

\begin{figure}[tb]
    \centering
    \includegraphics[width=1.\columnwidth]{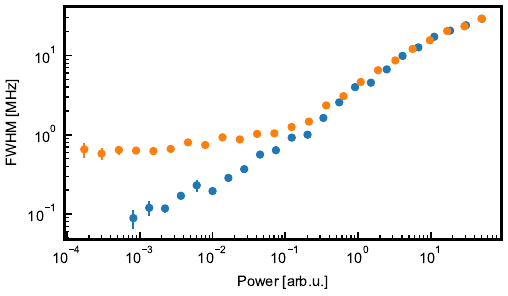}
    \caption{\label{fig:HomLW_AOM_SWEEP}
    \textbf{Effect of persistent spectral hole burning on the homogeneous linewidth.} The homogeneous linewidth is measured using the scheme described above. In the first curve (blue), the laser frequency is changed randomly after each iteration of the transient-hole-burning sequence. If the laser frequency is not shifted (orange), broader lines are obtained at low laser powers, which is attributed to persistent spectral hole burning.}
\end{figure}

To investigate this effect in more detail, we perform two measurements of the homogeneous linewidth on the CVD samples, one with the random frequency change, and one without. The result is shown in Fig.~\ref{fig:HomLW_AOM_SWEEP}. One observes an increased linewidth when the laser is repeatedly applied at the same frequency, which is a signature of persistent spectral hole burning. The width of the persistent hole, around $\SI{1}{\mega\hertz}$ in the shown measurement, depends on the experimental parameters of the sequence, including the temperature, repetition rate, and excitation pulse power and duration.

\section{Fit of the temperature dependence of the homogeneous linewidth}  \label{app:HomLinewidthTemperature}

The increase of the homogeneous linewidth with temperature (shown in Fig.~\ref{fig:Temperature dependence}) is caused by the reduction of the spin lifetime via direct, Raman or Orbach relaxation \cite{liu_spectroscopic_2005}. At zero magnetic field, the direct process is negligible. For $k_B \Theta_D < \Delta$, the Orbach process is expected to dominate over the Raman process \cite{hubner_simple_1996}. Here, $\Theta_D$ is the Debye temperature, $k_B$ Boltzmann's constant, and $\Delta$ energy splitting to the nearest crystal field level in either the optical ground or excited state. As this condition is fulfilled for both sites A and B, we fit our data to the following function: 
\begin{equation}
    \Gamma_{\text{orb}}(T) = A_{\text{orb}} \sum_{\Delta_{ES}, \Delta_{GS}} \frac{ \Delta^3}{\exp\left( {\frac{\Delta}{T}} \right) - 1} + \Gamma_{\text{offset}}
\end{equation}
Here, the sum is over all CF levels in the ground state ($GS$) and excited state ($ES$), and we assume that their Orbach coefficient $A_{\text{orb}}$ is identical. In this way, we determine $A_{\text{orb,A}}=\SI{0.17(6)}{\second^{-1}\kelvin^{-3}}$ and  $A_{\text{orb,B}}=\SI{0.15(6)}{\second^{-1}\kelvin^{-3}}$ for site A and B, respectively. $\Gamma_{\text{offset}}$ is introduced as a free fit parameter representing the power broadened homogeneous linewidth for temperatures below the onset of the Orbach relaxation.

\section{Integration efficiency} \label{app:IntegrationEfficiency}
The samples used in this work exhibit a low dopant density in order to avoid erbium precipitation as well as reabsorption and amplified spontaneous emission that could lead to errors in the lifetime measurement. Together with back-reflections in the fiber setup that lead to oscillations of the reflected intensity, this low concentration has hindered absorption measurements. Thus, the fraction of dopants that are integrated at the newly-observed sites is estimated from the fluorescence signal.

To this end, the number of detected photons per excitation pulse is determined and divided by the independently measured fiber-coupling efficiency and detection efficiency in order to get the number of photons emitted into the waveguide. This is then compared to the maximum number of dopants that can be excited. The latter is obtained by multiplying the total number of dopants (known from the implantation dose and the waveguide area) with the independently measured ratio of the homogeneous and inhomogeneous linewidths.

Note that each of these quantities has a considerable error, such that the technique only allows deriving the order of magnitude of the integration efficiency. In addition, three important effects are ignored in its calculation: First, the waveguide loss and potential mirror imperfections. Second, only a fraction of the dopants will be excited by the laser field with the assumed power-broadened linewidth - those that are close to the center of the mode, where the field is largest. Finally, only a fraction of the photons will be emitted into the fiber-coupled waveguide mode. As discussed in Appendix \ref{app:LDOS}, this depends on the emitter position and orientation.

Thus, the actual integration efficiency will be considerably higher than the calculated lower bound, which is between $0.05\,\%$ and $1\,\%$, with significant fluctuations between different samples even of the same type. Considering the above-mentioned effects, we thus estimate that the integration efficiency is $\gtrapprox 1\,\%$. Future studies may aim at a precise determination and increase of the integration efficiency by optimizing the starting material and implantation procedure.

\begin{figure*}[tb]
    \centering
    \includegraphics[width=2.\columnwidth]{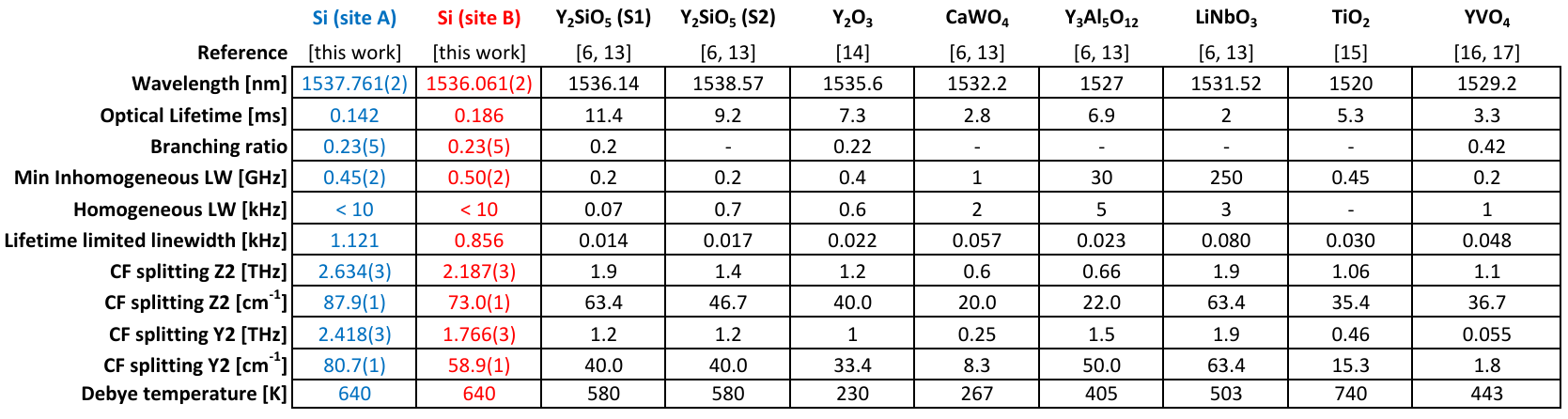}
    \caption{\label{tab:HostComparison} Comparison of the key parameters of erbium dopants in various hosts. The sites A and B studied in this work exhibit the largest CF splittings and the fastest radiative decay, offering unique promise for quantum applications.}
\end{figure*}

\section{Comparison to other host materials}  \label{app:OtherHostMaterials}

A comparison of the most prominent host materials \cite{liu_spectroscopic_2005, thiel_rare-earth-doped_2011, fukumori_subkilohertz_2020, phenicie_narrow_2019, xie_characterization_2021, li_optical_2020} for erbium dopants is given in Fig.~\ref{tab:HostComparison}. Earlier studies in silicon \cite{przybylinska_optically_1996, kenyon_erbium_2005, vinh_photonic_2009, weiss_erbium_2021} are not included because of the large number of sites and the lack of data on most of them. An exception is the work of Weiss et al. \cite{weiss_erbium_2021} that found inhomogeneous lines of $\sim \SI{1}{\giga\hertz}$, transient spectral hole widths of $\sim 4\cdot10^{4}\,\si{\kilo\hertz}$, and lifetimes of $\sim \SI{1}{\milli\second}$. Recent measurements by Berkman et al. \cite{berkman_sub-megahertz_2021} provide a larger survey, including sites with sub-MHz homogeneous linewidth.

\bibliographystyle{naturemagsr.bst}
\bibliography{bibliography.bib}


\end{document}